\documentclass[twocolumn]{aastex63}

\usepackage[T1]{fontenc}
\usepackage{microtype}
\usepackage{afterpage,amsmath,amssymb,graphicx,natbib, color}
\usepackage{hyperref}
\usepackage[english]{babel}
\usepackage{blindtext}
\usepackage{csquotes}
\usepackage[utf8]{inputenc}
\usepackage{enumitem}
\usepackage{graphicx}
\usepackage{hyperref}
\usepackage{mathtools}
\usepackage{comment}
\usepackage{amsmath}
\usepackage{ amssymb }
\usepackage{lipsum} 
\usepackage{float}
\usepackage{cleveref}
\usepackage[caption=false]{subfig}
\usepackage{placeins}
\usepackage{xcolor}
\usepackage{multirow}
\usepackage{float}
\usepackage{makecell}

\usepackage{amsmath}

\newcommand{\be}{\begin{equation}}
\newcommand{\ee}{\end{equation}}
\newcommand{\bfig}{\begin{figure}}
\newcommand{\efig}{\end{figure}}

\begin{document}
\title{Investigating the Reionization Epoch through 21\,cm and Line Intensity Mapping Experiments}

\correspondingauthor{Anirban Roy}
\email{ar8816@nyu.edu}

\author{Anirban Roy}
\affiliation{Department of Physics, New York University, 726 Broadway, New York, NY, 10003, USA}  
\affiliation{Center for Computational Astrophysics, Flatiron Institute, New York, NY 10010, USA}

\author{Anthony Pullen}
\affiliation{Department of Physics, New York University, 726 Broadway, New York, NY, 10003, USA}  
\affiliation{Center for Computational Astrophysics, Flatiron Institute, New York, NY 10010, USA}

\author{Patrick C. Breysse}
\affiliation{Department of Physics, New York University, 726 Broadway, New York, NY, 10003, USA}
\affiliation{Department of Physics, Southern Methodist University, Dallas, TX 75275, USA}

\author{Rachel S. Somerville}
\affiliation{Center for Computational Astrophysics, Flatiron Institute, New York, NY 10010, USA}


\begin{abstract}
The epoch of reionization (EoR), marking the Universe's transition from a neutral to ionized state, represents a pivotal phase for understanding the formation of the first stars and galaxies. Intensity mapping of atomic and molecular lines, such as $[\mathrm{CII}]$ and CO J-ladder transitions, across a broad redshift range is a powerful tool for investigating star formation history, metallicity, the distribution of gas and dust, and the physical conditions within galaxies. Additionally, 21\,cm line intensity mapping directly probes the neutral hydrogen content in the intergalactic medium, offering a unique window into the timing and morphology of reionization. In this study, we explore the cross-correlation between the 21\,cm signal and multi-line intensity mapping (LIM) to forecast their detectability for next-generation experiments. Our analysis emphasizes the complementary potential of these techniques to constrain parameters such as the minimum mass of ionizing sources and the ionization fraction $x_e$. Cross-correlations with LIM also enable constraints on physical properties like metal enrichment and the relationship between star formation rates and multi-line luminosities. Using mock observations from Square Kilometre Array (SKA)-low 21\,cm and Fred Young Submillimeter Telescope (FYST)-like LIM experiments, we find that the $[\mathrm{CII}]$--21\,cm cross-correlation can constrain reionization history by measuring $x_e$ across multiple redshift bins with significance levels ranging from 9 to 40$\sigma$. We extend our analysis to CO transitions, showing that the CO(1-0)--21\,cm cross-correlation provides competitive constraints on reionization parameters. The synergies explored here will enable robust constraints on both reionization and LIM parameters, maximizing the scientific return of current and next-generation intensity mapping experiments.
\end{abstract}

\section{Introduction}\label{sec:introduction}

Atomic and molecular emission lines from galaxies serve as crucial tracers of the complex physical processes underlying galaxy formation and evolution  \citep{Visbal2010, Visbal2011, Bernal2022-sfr}. Line intensity mapping, which captures the cumulative emission from these lines across a large three-dimensional cosmic volume, offers a unique opportunity to probe several poorly understood epochs in the history of the Universe \citep{Kovetz2017LIM_report}. This includes the formation of the first-generation stars, the peak of cosmic star formation rate density around $z \sim 1.9$, and the subsequent rapid decline, dust properties, heating and cooling processes in the interstellar medium, among other aspects \citep{Limfast2, Zhang-galaxy-property, Bernal2022-sfr, Zhou-lim-sfr}. By analyzing multiple emission lines from a given region of the sky, we can investigate the metal enrichment of galaxies and their environments, as well as their evolution within the cosmic overdensity field, providing critical insights into the interplay between baryonic physics and large-scale structure \citep{Suginohara1998, Righi2008b, Lidz2011_CO, Carilli2011, Fonseca:2016, Gong2017, Kovetz2017LIM_report, Chung2018CII, PadmanabhanCO, Padmanabhan_CII, Dumitru2018, Chung2018CO, Kannan:2021ucy, Murmu:2021ljb, Karoumpis2021, Limfast2, slick-garcia}.

The 21\,cm hyperfine transition of neutral hydrogen traces the spin temperature and ionization state of the intergalactic medium, with emission strength depending on the gas density, temperature, and local radiation field during the pre- and post-reionization epochs. The [CII] 158 $\mu$m fine-structure line primarily originates from photo-dissociation regions in star-forming galaxies, where far-ultraviolet radiation from massive stars creates partially ionized carbon in dense molecular clouds and HII region boundaries. CO rotational transitions probe the cold, dense molecular gas within galaxies, with lower-J transitions (CO(1-0), CO(2-1)) tracing the bulk molecular reservoir while higher-J lines (CO(3-2), CO(4-3)) emerge from warmer, denser regions closer to star formation sites \citep{2020A&A...643A.141M, Kamenetzky2016}. These complementary emission mechanisms allow multi-tracer intensity mapping to probe different phases of the interstellar and intergalactic medium simultaneously.

Intensity mapping of emission lines at high redshift represents a powerful observational approach for investigating the cosmic transition from a predominantly neutral hydrogen medium to a fully ionized state during the Epoch of Reionization (EoR). Current theoretical frameworks combined with indirect observational constraints—including Ly$\alpha$ forest analyses, proximity zone studies around high-redshift quasars, and Thomson scattering optical depth measurements from cosmic microwave background observations—indicate that the reionization process completed within the redshift interval $z \sim 5-6$ \citep{Becker2015, kulkarni2019l-reio, P18:main, Nasir:2019iop, Bosman:2021oom, Davies:2023gin, Gaikwad:2023ubo}. 

A primary unresolved question concerns the identification of the dominant ionizing sources responsible for cosmic reionization. Pre-JWST investigations suggested that star-forming galaxies served as the primary drivers of reionization, given the observed paucity of luminous quasars at $z > 6$. However, recent JWST discoveries have revealed an unexpectedly abundant population of moderate-luminosity active galactic nuclei (AGN) spanning $4 \lesssim z \lesssim 13$, necessitating a reassessment of the relative contributions from galactic stellar populations versus AGN in driving the reionization process \citep{Madau-2024-agn-vs-galaxy}. Additionally, the specific properties of reionizing galaxy populations remain poorly constrained. Key uncertainties include the luminosity function of contributing galaxies, particularly at the faint end where number densities are highest. Furthermore, the efficiency of ionizing photon escape from galactic environments—quantified through the escape fraction—represents a critical but poorly understood parameter. Recent analyses suggest that JWST-detected galaxies may produce ionizing photon budgets exceeding those required for the observed reionization timeline, implying potentially lower escape fractions than previously assumed \citep{Papovich-2025-reionization}. This apparent tension between high intrinsic ionizing emissivity and the inferred reionization redshift highlights the complex interplay between source properties and environmental factors governing photon escape. Uncertainties in identifying reionization sources, their luminosity functions, and escape fractions impact reionization process, driving further observational and theoretical studies of this crucial event in the cosmic timeline.

Intensity mapping with [CII] and CO J-ladder transitions has been proposed to study the EoR due to their brightness and their frequency overlap with current experiments, such as EoR-Spec at FYST \footnote{\href{https://www.ccatobservatory.org/}{https://www.ccatobservatory.org/}} \citep{CCAT-prime2021}, COMAP \footnote{\href{https://comap.caltech.edu/}{https://comap.caltech.edu/}} \citep{COmap-science-2021}, CONCERTO \footnote{\href{https://mission.lam.fr/concerto/}{https://mission.lam.fr/concerto/}} \citep{CONCERTO-science-2020}, and TIME \citep{Time-science-2014}. Other LIM experiments, including SPT-SLIM \citep{Karkare:2021ryi}, EXCLAIM  \citep{EXCLAIM-2020}, and SPHEREX\footnote{\href{https://spherex.caltech.edu/}{https://spherex.caltech.edu/}} \citep{SPHEREx-science-paper2018}, will focus on probing the post-reionization epochs using multiple spectral lines. These efforts will provide critical insights into the subsequent structure formation following reionization. Assimilating intensity maps of multiple lines covering the same redshift can significantly reduce systematic effects and improve the fidelity of parameter constraints at that redshift \citep{Roy-Lim-LLX}. For studying the reionization process, both the cross-correlation between two lines at $z \gtrsim 5.5$ and the cross-correlation between multi-line intensity maps and external tracers, such as galaxy distributions, would provide a comprehensive picture of the reionization process.

A special case of multi-line intensity mapping comes from 21\,cm emission from neutral hydrogen during the EoR. The detection of the 21\,cm line has been a central focus in reionization studies over the past few decades. While a direct detection of the 21\,cm signal from the EoR remains elusive, it holds significant promise for unraveling key scientific questions related to the reionization epoch. Current experiments, such as Low-Frequency Array (LOFAR) \citep{lofar-science-book}, Murchison Widefield Array (MWA) \citep{2016ApJ...819....8P}, and Hydrogen Epoch of Reionization Array (HERA) \citep{Hera}, along with future facilities like the Square Kilometre Array (SKA) \citep{ska-science-book}, are expected to substantially enhance our understanding of reionization through observations of the 21\,cm line across multiple redshifts, spanning the cosmic dawn and the EoR. The primary challenge in 21\,cm observations arises from the overwhelming level of foreground contamination, which is four to five orders of magnitude stronger than the expected signal. Despite this, the power spectrum of the 21\,cm signal provides a wealth of information, being sensitive to the minimum halo mass of ionizing sources, the size distribution of ionized (HII) regions, and the ionization fraction at a given redshift. 

The 21\,cm signal and multi-line intensity mapping (MLIM)  are highly complementary probes of the EoR, although both face significant challenges in detecting the signals. The 21\,cm signal traces the neutral hydrogen regions, while MLIM targets the ionized regions surrounding the sources of reionization. As a result, these two observables are expected to be anti-correlated on large scales. Furthermore, this cross-correlation evolves with redshift, providing a means to track the progression of reionization and its sensitivity to the parameters governing the reionization process. Previous investigations have explored the potential of [CII]-21\,cm cross-correlation analyses for constraining reionization parameters. \citet{Gong2012} demonstrated the feasibility of joint [CII]  and 21\,cm observations for accessing reionization physics while mitigating low-redshift CO line contamination, though their error analysis did not incorporate the impact of foreground contamination on binned cross power spectrum measurements. Subsequent work by \citet{Dumitru2018} extended this framework to constrain key reionization parameters, including the minimum halo mass threshold for IGM ionization and the ionizing photon production efficiency per unit halo mass, across various [CII]  and 21\,cm survey configurations. More recently, \citet{fronenberg2024forecasts} implemented a full end-to-end simulation pipeline incorporating foreground residuals and instrumental systematic effects, assessing signal detectability under Gaussian and non-Gaussian statistical assumptions for [CII]-21\,cm cross-power spectrum analyses, with observational predictions for HERA and CCAT-like survey designs.

In this work, we provide updated assessments of cross-correlation signals across different LIM modeling frameworks. Our analysis incorporates observational systematics including foreground contamination in 21\,cm measurements, instrumental beam responses, and interloper line confusion in LIM surveys. We adopt physically motivated reionization scenarios with ionization histories consistent with existing observational constraints. We evaluate parameter inference capabilities under conservative, optimistic, and intermediate signal strength assumptions for LIM tracers. Furthermore, we generalize this methodology to analyze 21\,cm cross-correlations with additional low-frequency LIM probes, encompassing CO(1-0) through CO(4-3) rotational transitions, and assess their contributions to parameter estimation precision. We also investigate whether multi-line combinations enable tomographic reconstruction of the cosmic ionization fraction evolution. This approach provides constraints on both reionization morphology and line intensity mapping properties through the synergistic exploitation of 21\,cm and LIM observational datasets.

This paper is organized as follows: In Section \ref{sec:theory}, we introduce the theoretical framework for the 21\,cm signal from neutral hydrogen and the multi-line intensities from galaxies. Section \ref{sec:sims} outlines our simulation-based approach for generating LIM maps for various atomic and molecular lines using the \texttt{LIMpy} package \citep{limpy}. We describe the application of the excursion-set approach to create ionized bubbles during the EoR and detail the steps involved in this process \citep{Park:2021eux, Kulkarni:2017qwu, Mesinger:2007pd}. In the following section, we investigate the cross-correlation between the 21\,cm and multi-line intensity signals at different redshifts during the EoR, exploring the evolution of these signals. In Section \ref{sec:exp_forecast}, we describe the foreground models for 21\,cm and LIM experiments. Section \ref{sec:forecast} forecasts the detectability of these signals for the experiments considered in this study and provides a detailed analysis of constraints on parameters that offer insights into both LIM and reionization. Finally, we summarize our findings and conclusions in Section \ref{sec:conclusion}. Throughout this work, we adopt a flat $\Lambda$CDM cosmological model, constrained by parameters from the Planck TT, TE, EE+lowE+lensing analysis \citep{P18:main}. It is important to note that while 21\,cm signals can be considered a form of line intensity mapping, the term ``line intensity mapping'' in this context specifically refers to atomic and molecular lines, excluding the 21\,cm signal.

\section{Theoretical background}\label{sec:theory}
In this section, we review the ionization history of the Universe and provide the theoretical background of the 21\,cm signal as a probe of the neutral hydrogen during the EoR. Additionally, we provide a brief overview of the LIM techniques and their significance in studying the ionized regions and large-scale structure of the Universe.

\begin{figure*}[t]
\includegraphics[width=\textwidth]{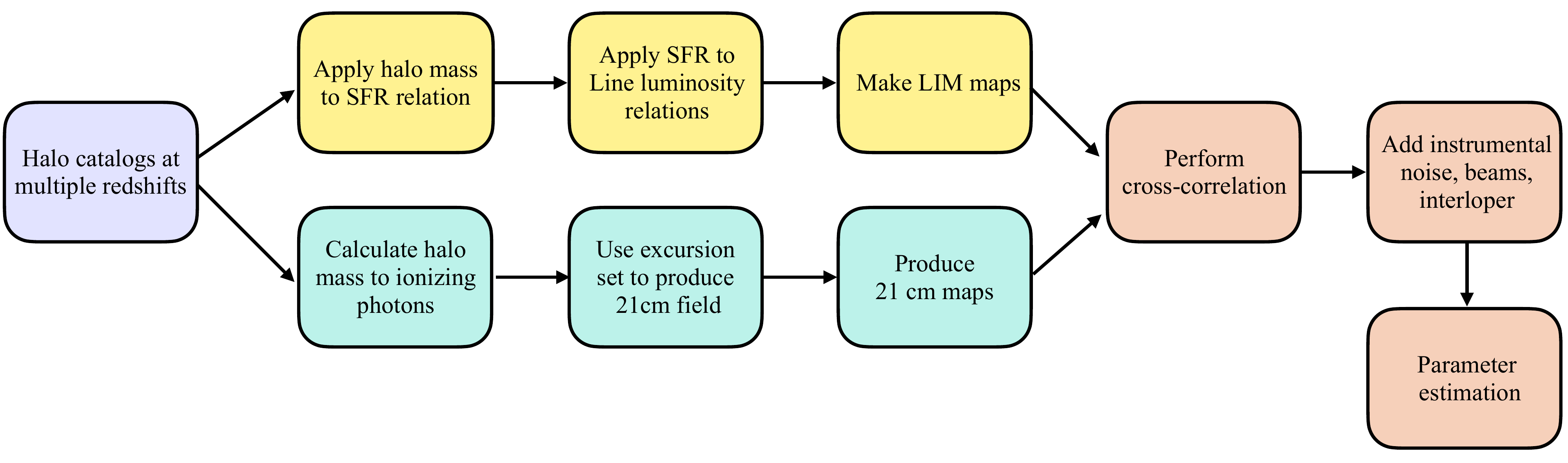} 
\caption{Schematic workflow for [CII]-21\,cm and COs-21\,cm cross-correlations analysis. We model [CII]/CO emissions using halo mass-to-SFR and SFR-to-luminosity scaling relations, and 21cm signals via excursion set formalism applied to ionizing photon production. Cross-correlation of the resulting intensity maps enables power spectrum forecasts. Finally, we incorporate instrumental noise, interlopers, foregrounds, and other effects to forecast the detectability of these cross-correlations and determine the constraints on the parameters of interest.}
\label{fig:flowchart}
\end{figure*}

\subsection{Ionization History}\label{subsec:reionization_history}

\begin{figure}[h]
\includegraphics[width=0.5\textwidth]{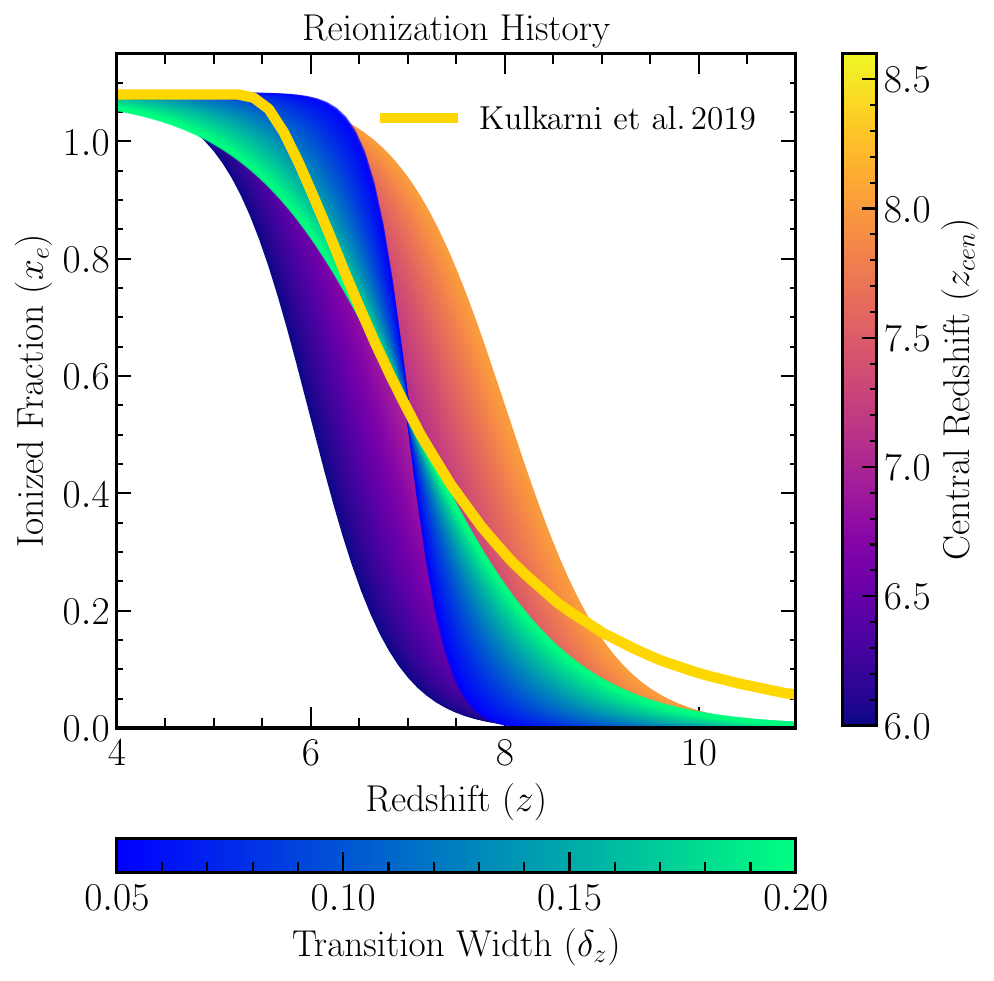} 
\caption{Evolution of the free-electron fraction for a \textit{tanh} reionization history. We illustrate the change in the ionization fraction resulting from variations in the central redshift of reionization and the width of reionization using two adjacent colorbars. The reionization history derived from the radiative transfer simulations by \citet{kulkarni2019l-reio} is overplotted for comparison.}
\label{fig:xe}
\end{figure}

For simplicity, a tanh-like reionization history with two free parameters is
commonly adopted, following \citet{Lewis:2008wr},
\begin{equation}
x_e(z) = \frac{f}{2}
\left[
1 + \tanh\left( \frac{y_{\rm re} - y(z)}{\Delta_y} \right)
\right] .
\end{equation}

In this expression, $x_e(z)$ denotes the free-electron fraction at redshift $z$,
defined as the ratio of the number density of free electrons to the number
density of hydrogen atoms, $x_e \equiv n_e / n_{\rm H}$. In addition to hydrogen,
an extra $\simeq 8\%$ contribution to the free-electron density arises from the
first ionization of helium. This contribution is absorbed into the parameter
$f$, which sets the maximum electron fraction. We therefore take
$f = 1 + f_{\rm He} \simeq 1.08$, corresponding to fully ionized hydrogen and
singly ionized helium. The parameter $y_{\rm re}$ defines the midpoint of the reionization transition,
corresponding to the redshift at which the ionized fraction reaches half of its
maximum value, while $\Delta_y$ controls the width of the transition, describing
the redshift interval over which the ionization fraction increases from
approximately 10\% to 90\%.

The term \(y(z)\) is defined as $y(z) = (1 + z)^{3/2}$. The relationship between \(\Delta_y\) and the width of reionization in redshift space is given by $\Delta_y = 1.5 \times (1 + z)^{1/2} \Delta_z\,$ where \(\Delta_z\) represents the redshift width of the reionization transition, defined as the difference between the redshift where the ionization fraction is 90\% and the redshift where it is 10\%. This relationship arises because \(y(z)\) scales with \((1 + z)^{3/2}\), leading to a proportional change in \(y\) over the reionization period, which is captured by \(\Delta_z\). 

We illustrate various scenarios for the ionization state of the Universe by showing the evolution of $x_e$ in Figure \ref{fig:xe}. In the first case, we keep the redshift of reionization, $z_{\rm re}$, fixed at 7.0 while varying $\Delta z$ from 0.05 to 0.20. In the second case, we fix $\Delta z$ at 0.12 and vary $z_{\rm re}$ from 6 to 8. For comparison, we include the reionization history from \citet{kulkarni2019l-reio}, which is derived from high-dynamic-range radiative transfer simulations, denoted by ``Kulkarni19''. This reionization history is also consistent with constraints from the spectra of the two highest-redshift quasars and the model-independent constraints from the incidence of dark pixels in Ly$\alpha$ forests, Quasar near-zone observations, and the optical depth constraint from Planck data \citep{kulkarni2019l-reio}. Unless otherwise mentioned, we use the ``Kulkarni19" reionization history as the fiducial model for this work.

\begin{figure*}[t]
\includegraphics[width=\textwidth]{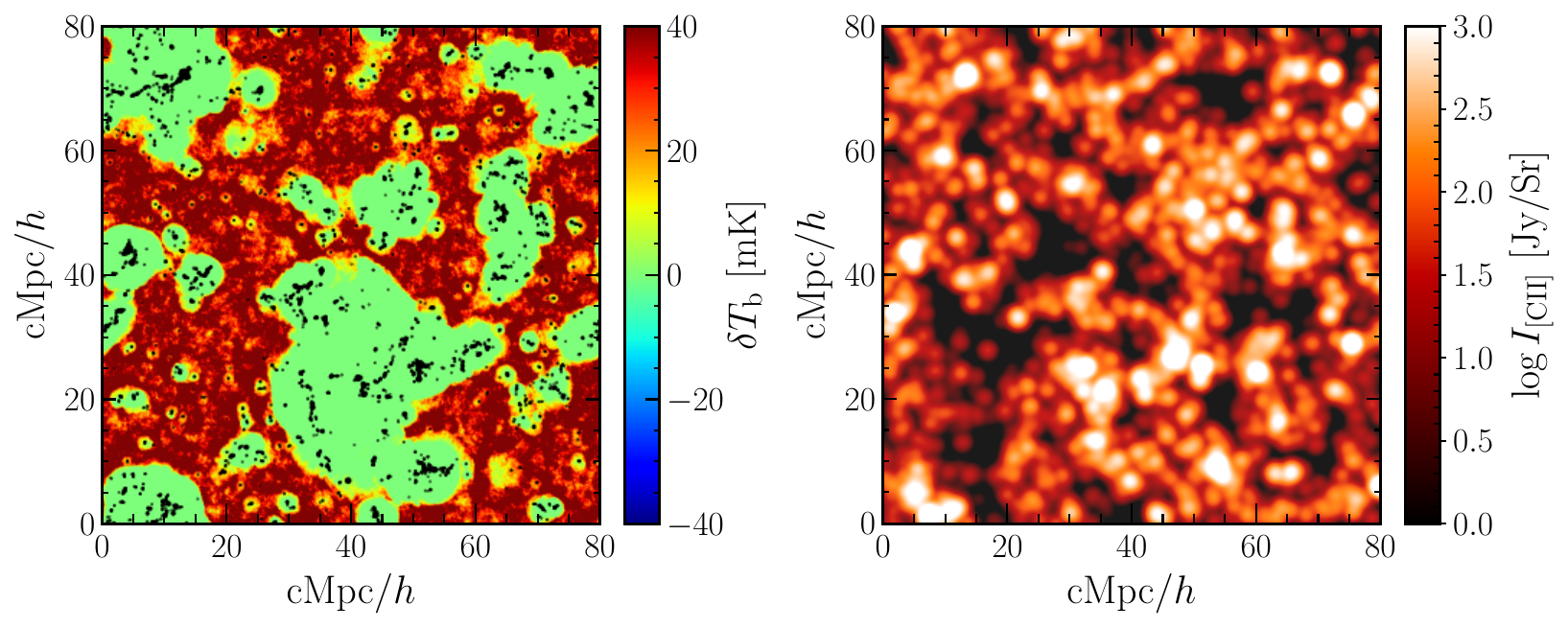} 
\caption{A snapshot of the 21\,cm brightness temperature fluctuations and [CII] intensity emission at \( z \sim 7.3 \). The black dots on the 21\,cm panel represent the distribution of dark matter halos. For visibility purposes, we apply a beam convolution with \( \theta = 1\,\mathrm{arcmin} \) to the [CII] intensity map.}
\label{fig:slice_plot}
\end{figure*}

\subsection{21\,cm field}\label{subsec:21cm_theory}
The 21\,cm brightness temperature fluctuation, $\delta T_b$, represents the difference between the redshifted 21 cm brightness temperature $T_b(z)$ and the redshifted CMB sky-averaged temperature $T_{\text{CMB}}(z)$. The latter is given by the relation $T_{\text{CMB}}(z) = T_0(1+z)$, where $T_0 \approx 2.73 \, \text{K}$ is the CMB temperature at redshift $z = 0$. For small optical depths, the fluctuation $\delta T_b$ can be approximated as \citep{Furlanetto2006}:
\begin{equation}\label{eq:deltatb}
\delta T_b \approx \frac{T_S - T_{\text{CMB}}}{1+z} \tau_{21},
\end{equation}
where $T_S$ is the spin temperature of neutral hydrogen, $T_{\text{CMB}}$ is the CMB temperature at the given redshift, and $\tau_{21}$ is the optical depth of the 21\,cm line.

During the epoch of reionization, both the neutral hydrogen fraction, $x_H(z)$, and the optical depth $\tau_{21}$ vary. These variations affect the 21\,cm brightness temperature fluctuation, $\delta T_b$. Therefore, equation \eqref{eq:deltatb} can be reformulated as \citep{Meerburg2013}:
\begin{equation}
\begin{split}
\delta T_b(z) = & \, 27~\mathrm{mK} \, (1 + \delta_b) \, x_H(z) \\
& \times \left[ \frac{T_S - T_{\text{CMB}}}{T_S} \right] \left[ \frac{1 - Y_P}{1 - 0.248} \right] \\
& \times \left[ \frac{\Omega_b}{0.044} \right] \left[ \left( \frac{0.27}{\Omega_m} \right) \left( \frac{1+z}{10} \right) \right]^{1/2},
\end{split}
\end{equation}
where $\delta_b$ is the baryon overdensity, $Y_P$ is the primordial helium fraction, and $\Omega_b$ and $\Omega_m$ are the baryon and matter density parameters, respectively. The neutral hydrogen fraction $x_H(z)$ and the baryon overdensity $\delta_b$ depend on the reionization history of the universe, which influences the 21\,cm signal. The free-electron fraction is defined as
\(x_e \equiv n_e/n_{\rm H}\), where
\(n_{\rm H} = n_b(1-Y_P)\) is the hydrogen number density.
In the absence of partially ionized regions,
the neutral hydrogen fraction is related by
\(x_H = 1 - x_e\).

\subsection{Multi-line intensity mapping}\label{subsec:lim_theory}
Prominent emission lines important for the current and future LIM experiments include the [CII] 158 $\mu$m line, CO rotational lines, and the [OIII] 88 $\mu$m line, each offering unique information about the galaxy's properties. The [CII] 158 $\mu$m line, arising from a fine structure transition of ionized carbon, is a crucial tracer of the ISM in galaxies. It is strongly correlated with the SFR and is particularly valuable for studying high-redshift galaxies. The luminosity of the [CII] line can be expressed as a linear function in the log-scale of SFR \citep{limpy, Schaerer_2020}:

\begin{equation}
\log L_{\text{CII}} = a_{\text{CII}} + b_{\text{CII}} \log \text{(SFR)},
\end{equation}

where $L_{\text{CII}}$ denotes the [CII] luminosity, and $a_{\text{CII}}$ and $b_{\text{CII}}$ are empirically determined coefficients. This relationship allows for the conversion of observed [CII] intensities into estimates of star formation activity for the set of values of $a_{\text{CII}}$ and $b_{\text{CII}}$.

The CO emission is often used to estimate the amount of molecular gas, which is essential for understanding the kinematics and cooling properties in galaxies. The luminosity of CO lines, such as CO(1-0) and CO(2-1), can be similarly parameterized \citep{Kamenetzky2016}:

\begin{equation}
\log L_{\text{CO}(J)} = a_{\text{CO}(J)} + b_{\text{CO}(J)} \log \text{SFR},
\end{equation}

where $L_{\text{CO}(J)}$ represents the luminosity of the CO transition from rotational level $J$ to $(J-1)$, and $a_{\text{CO}(J)}$ and $b_{\text{CO}(J)}$ are coefficients that reflect the empirical calibration of the line emission. CO lines provide complementary information to [CII] measurements by probing the gas density and temperature.

The [CII]-SFR power law relationship arises because [CII] emission originates from photo-dissociation regions where UV photons from young, massive stars ionize carbon atoms in the neutral ISM. Since star formation rate directly determines the UV photon budget through the number and luminosity of newly formed stars, [CII] luminosity scales predictably with the energy input from star formation. The power-law form reflects the underlying physical scaling between stellar UV radiation field strength and the cooling efficiency of [CII] emission in the multi-phase interstellar medium. The observational robustness was further strengthened by combining ALPINE data and creating a homogenized sample of $\sim$150 galaxies spanning $z \sim 4-8$, which showed consistent behavior across this redshift range \citep{Schaerer_2020}. Additionally, the measured $L_{\rm [CII]}/L_{\rm IR}$ ratios of $\sim(1-3) \times 10^{-3}$ for ALPINE sources matched those of lower-redshift galaxies, supporting the physical consistency of the [CII]-SFR correlation over cosmic time.

\citet{Kamenetzky2016} construct CO spectral line energy distributions (SLEDs)
by measuring luminosity ratios of mid- to high-$J$ CO rotational transitions
relative to the CO($J{=}1{\rightarrow}0$) ground-state line, using Herschel
SPIRE FTS observations to characterize molecular gas excitation conditions.
Line intensities are inferred using a custom Bayesian analysis framework
designed to account for correlated instrumental noise, and are calibrated
against low-$J$ CO measurements from ground-based observations. The resulting
SLEDs provide empirical power-law relations between the far-infrared
luminosity $L_{\rm FIR}$ (or $L_{\rm IR}$) and the CO brightness--temperature
luminosity $L'_{\rm CO}$ for each rotational transition.

We convert the commonly used
brightness--temperature luminosity $L'_{\rm CO}$ into the corresponding
line luminosity $L_{\rm CO}$ using the standard relation
\citep[e.g.,][]{Li2016}
\begin{equation}
\frac{L_{\rm CO}}{L_\odot}
=
4.9 \times 10^{-5}
\left(
\frac{\nu_{\rm CO,rest}}{115.27\,\mathrm{GHz}}
\right)^3
\left(
\frac{L'_{\rm CO}}{\mathrm{K\,km\,s^{-1}\,pc^2}}
\right)\, ,
\end{equation}
which introduces an explicit dependence on the rest-frame frequency of each
CO transition. To connect CO emission to star formation, we convert the star formation rate
to infrared luminosity using the scaling relation
$ \mathrm{SFR} = 1 \times 10^{-10} L_{\rm IR}$
\citep{Carilli2011}. Combining this relation with the empirical power-law
relations between $L'_{\rm CO}$ and $L_{\rm IR}$ from
\citet{Kamenetzky2016}, and the above conversion to $L_{\rm CO}$, yields a
power-law relation between CO line luminosity and SFR. The amplitude and slope
of this relation are treated as free parameters in our model and are
constrained using the observational data considered in this work.

In our modeling, the parameters $a_{\rm off}$ and $b_{\rm off}$ describe the
$[\mathrm{CII}]$ luminosity--SFR relation. In general, the luminosity--SFR
relation for each CO rotational transition can be characterized by its own
independent pair of parameters $(a_{\rm CO}, b_{\rm CO})$, reflecting the
highly uncertain and line-dependent physical conditions of molecular gas in
high-redshift galaxies. These parameters are expected to vary across
transitions and may also exhibit strong and uncertain redshift dependence,
which we neglect in this work.

In order to aid in comparing the sensitivities to different lines, we normalize the various slopes and amplitudes to the values of the [CII] line as 
\[
r_a \equiv \frac{a_{\rm CO}}{a_{\rm off}},
\qquad
r_b \equiv \frac{b_{\rm CO}}{b_{\rm off}}.
\]
We fix these ratios to fiducial values derived from empirical CO spectral line
energy distribution (SLED) measurements \citep[e.g.,][]{Kamenetzky2016}, using
CO(1-0) as a reference, and apply the same normalization across all CO
transitions considered in this work.

With this convention, the CO luminosity--SFR parameters are written as
\[
a_{\rm CO}^\ast = r_a\, a_{\rm off}^\ast,
\qquad
b_{\rm CO}^\ast = r_b\, b_{\rm off}^\ast,
\]
where $a_{\rm off}^\ast$ and $b_{\rm off}^\ast$ are the parameters constrained
by our forecasts. This serves as a convenient normalization
that enables a clear illustration of the relative sensitivities of different
CO transitions compared to $[\mathrm{CII}]$ within a common parameter
framework. A more general treatment allowing independent $(a_{\rm CO},
b_{\rm CO})$ for each line is left for future work.

\section{Simulations of 21\,cm and LIM}\label{sec:sims}
In this section, we describe how we generate the simulations of the 21\,cm signal and LIM signals. In the first step, we generate halo catalogs and the density fields using approximate N-body simulations. Afterwards, we use these outputs to calculate the 21\,cm signal based on different reionization histories and LIM signals for different models. Below, we describe the methods for generating the 21\,cm and LIM signals separately.

\begin{figure*}[t]
\includegraphics[width=\textwidth]{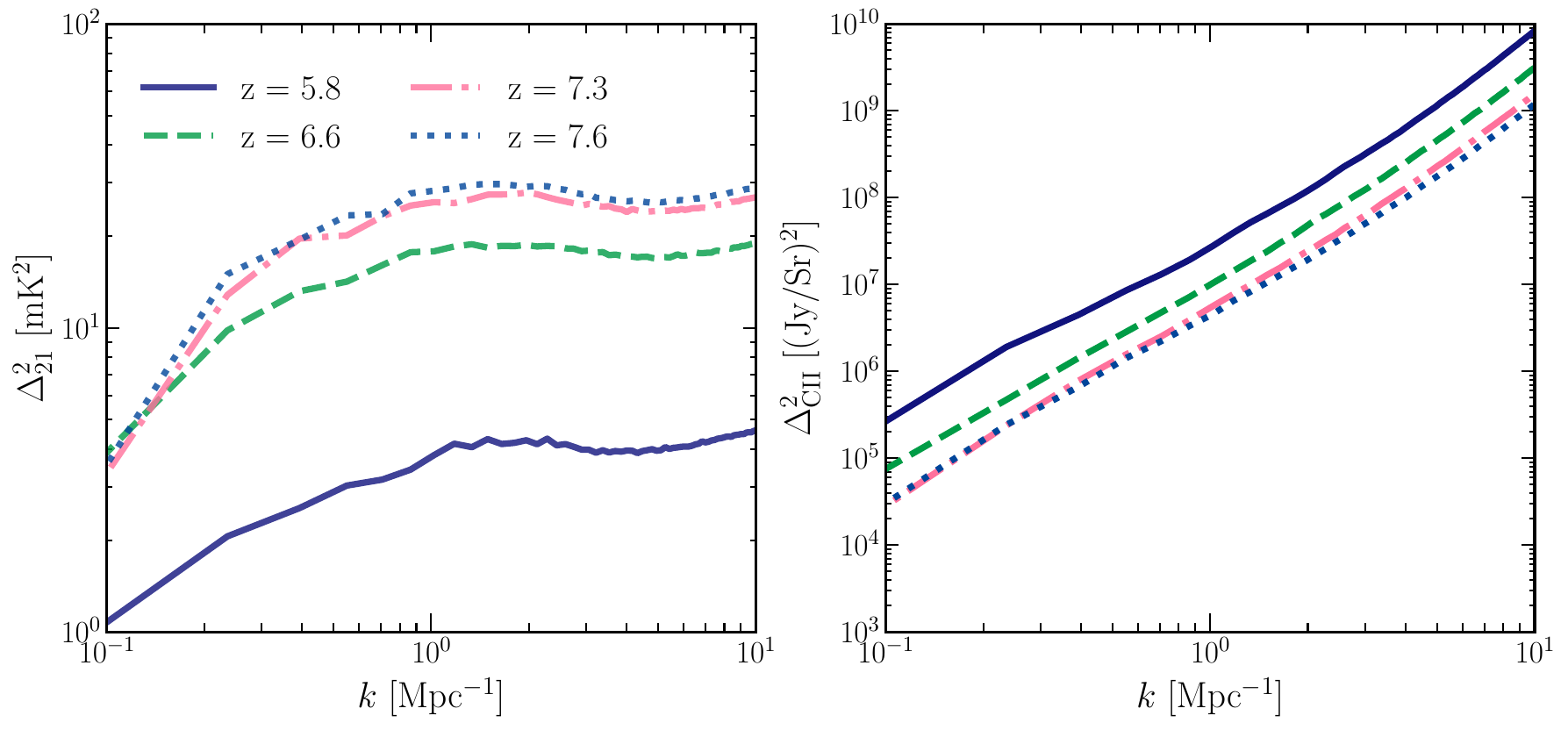} 
\caption{ Left: the auto power spectrum of the 21\,cm signal at redshifts 7.6, 7.3, 6.6, and 5.8 for ionization fractions of 0.35, 0.42, 0.63, and 0.91, respectively. The 21\,cm power spectrum decreases with decreasing redshift as the Universe becomes more ionized, reducing the abundance of neutral hydrogen. Right: the power spectrum of [CII] line emissions at the same redshifts as the 21 cm signal. The power spectrum increases with redshift as more sources produce high-energy photons that ionize the IGM. As we integrate over all sources, the amplitude increases. For the LIM calculation, we use the \textit{Schareer20} models to generate the LIM maps.}
\label{fig:ps_all}
\end{figure*}

\subsection{21cm maps}\label{subsec:21cm_sim}
To model the ionization field based on the density field and dark matter haloes, we employ the excursion-set formalism, positioning sources of UV photons within the haloes. Following the approach of \cite{2016MNRAS.463.2583K}, we assume that the total number of ionizing photons, \( N_\gamma \), produced by a halo is directly proportional to its mass \( M_{\rm halo} \) \citep{Choudhury:2014uba, Majumdar2012}. This relationship is expressed as:

\begin{equation}
  N_\gamma(M_{\rm halo}) = \frac{N^{\rm esc}_\gamma M_{\rm halo}}{m_p},
  \label{eqn:ngammam}
\end{equation}

where \( N_\gamma^{\rm esc} \) is a proportionality factor incorporating the escape fraction and \( m_p \) is the proton mass.

A grid cell located at position \(\vec{x}\) is deemed ionized if the following criterion is satisfied within a spherical region of radius \( R \) centered on the cell:

\begin{equation}
  \langle n_\gamma(\vec{x}) \rangle_R > \langle n_p(\vec{x}) \rangle_R (1 + \bar{N}_\mathrm{rec}),
  \label{eqn:exset1}
\end{equation}

Here, \(\langle \cdot \rangle_R\) denotes the average over the spherical region, \( n_p \) is the number density of hydrogen nuclei, and \( n_\gamma \) represents the ionizing photon density, calculated as \citep{Furlanetto2004, Mesinger2011}:

\begin{equation}
  n_\gamma = \int_{M_\mathrm{min}}^\infty dM \left. \frac{dN}{dM} \right\vert_{R} N_\gamma(M).
  \label{eqn:ngamma}
\end{equation}

In this expression, \( \frac{dN}{dM}\big\vert_{R} \) is the halo mass function within the spherical region, \( M_\mathrm{min} \) is the minimum halo mass capable of contributing ionizing photons. The global volume average of \( n_p \) is denoted by \( \bar{n}_p \). The condition in Eq.~\eqref{eqn:exset1} ensures that sufficient photons are generated to ionize all hydrogen atoms within the region while accounting for recombinations. This condition can be reformulated as:

\begin{equation}
  \zeta_\mathrm{eff} f(\mathbf{x}, R) \geq 1,
  \label{eqn:exset}
\end{equation}

where \( f \) represents the fraction of mass collapsed into haloes with \( M > M_\mathrm{min} \) within the spherical region centered on \(\vec{x}\). It is given by:

\begin{equation}
  f = \rho_m(R)^{-1} \int_{M_\mathrm{min}}^\infty dM \left. \frac{dN}{dM} \right\vert_{R} M,
  \label{eqn:fcoll}
\end{equation}

Here, \( \rho_m(R) \) is the average matter density within the sphere. The parameter \( \zeta_\mathrm{eff} \) quantifies the ionizing photon density in the IGM per hydrogen atom in stars, adjusted for recombinations, and is defined as:

\begin{equation}
  \zeta_\mathrm{eff} = \frac{N^\mathrm{esc}_\gamma}{1 - Y_\mathrm{He}} \left(1 + \bar{N}_\mathrm{rec} \right)^{-1},
\end{equation}

where \( Y_\mathrm{He} \) is the helium mass fraction. \( \zeta_\mathrm{eff} \) serves as the sole parameter governing the ionization field in this model.

Following this approach, we generate the simulation snapshots consisting of the ionized and neutral regions at different redshifts. When the ionization fraction is low, the ionized bubbles will start appearing at the high-mass halos as they produce a large number of UV photons to ionize the medium surrounding them. These ionized regions are also correlated with the density field. We then generate the grids of $x_e$ at different redshifts for the fixed ionization fraction corresponding to those specific redshifts. Afterwards, we use those density grids and ionization grids at a few redshifts in the EoR, and   calculate the 21\,cm signal. We use this grid of 21\,cm signals for the study of cross-correlations with the LIM signal. The semi-numerical framework employed in this study exhibits a lack of numerical convergence in ionization power spectrum estimates, due to non-conservation of photons \citep{Zhan2007, Choudhury-photon-noncon}. We nevertheless utilize this approach given its established calibration against radiative transfer simulations across mutually accessible scales, maintaining concordance with empirical measurements \citep{Majumdar2016, Roy2021-b-mode, Hutter2018}.

\subsection{Line intensity maps}\label{subsec:lim_sim}
We adopt a semi-empirical framework to generate CO intensity maps based on the
CO spectral line energy distribution model of
\citet{Kamenetzky2016}. For $[\mathrm{CII}]$ emission, we employ the model
proposed by \citet{Schaerer_2020}, which is calibrated using ALMA-ALPINE data
and is chosen for its simple and transparent parameterization. Throughout this
work, we treat these models as fiducial descriptions of the line--SFR
relations and neglect additional astrophysical scatter and stochastic effects
unless explicitly stated otherwise. This modeling framework has been implemented within the \texttt{LIMpy}
package \citep{limpy}, enabling efficient generation of line intensity maps
from externally provided halo catalogs. The modular structure of
\texttt{LIMpy} facilitates direct comparisons between different line
luminosity prescriptions and substantially reduces the computational cost of
map generation when halo catalogs from $N$-body simulations are available. By
adopting different SFR prescriptions, the framework can
be used to produce intensity maps for multiple emission lines relevant to this
study. In practice, we assign line luminosities to halos drawn from external
$N$-body simulations and project these onto intensity maps, allowing us to
generate both target signal maps and interloper line maps required for mock
observations. 

For constructing halo catalogs at target redshifts, we employ approximate $N$-body simulations using the \texttt{Nbodykit} software package \citep{Nbodykit}. This computational approach offers significant efficiency advantages for our Fisher forecast analysis, which requires generating multiple density realizations across varying reionization scenarios and physical parameter sets. Approximate methods for density and halo catalog generation provide statistically equivalent representations of the underlying density field, making them well-suited for reionization studies at intermediate scales ($0.05 \lesssim k \lesssim 1$ h Mpc$^{-1}$). Our simulation framework encompasses redshift slices spanning $z = 5.8$ to $7.6$, implemented within cubic volumes of $(80 \text{ Mpc}/h)^3$ with spatial resolution of approximately $0.156$ Mpc$/h$ achieved through $N_{\text{grid}} = 512^3$ grid sampling. The corresponding dark matter particle mass is $3.4 \times 10^8 \, M_{\odot}/h$. Halo identification employs a friends-of-friends (FoF) algorithm with linking length parameter $b = 0.2$, adopting virial mass definitions for halo mass assignments. This configuration enables reliable detection of halos with masses $M_{\text{halo}} \geq 10^{10} \, M_{\odot}/h$.

We use the same halo catalog for generating the 21\,cm and LIM signals for consistency. Using \texttt{LIMpy}, we can paint the line intensities on the halos based on several SFR models and line-to-luminosity-to-SFR models. For example, there are five built-in SFR models, such as Fonseca16 \citep{Fonseca:2016}, Silva15\citep{Silva:2015}, Tng100, Tng300 \citep{TNG-gal, TNG-gen}, and Behroozi19 \citep{Behroozi2019}. We suggest the reader check \citet{limpy} for details of these built-in models. After generating the LIM signal using a halo catalog for a specific model, we use that snapshot for post-processing. For example, \texttt{LIMpy} allows for the incorporation of frequency resolution ($\delta \nu$) along the redshift axis when creating the 3D line intensity maps. This means that an experiment will not be able to resolve the sources that fall within $\Delta z$ corresponding to $\delta \nu$. Additionally, the beam-convolution effect can be applied to the simulation grid of LIM signals for a beam size given by an experiment. In our case, we use the ``Behroozi19'' SFR model as our fiducial model for generating the LIM maps.

\subsection{LIM $\times$ 21\,cm signal }\label{sec:cross_signal}
We use the precomputed 21\,cm and LIM simulation grids to estimate their cross-correlations during the EoR. We present the slices of the fluctuations in the 21\,cm signal and [CII] intensities in Figure \ref{fig:slice_plot} at \( z \sim 7.3 \). The 21\,cm slice is projected over a 1.5\,cMpc/$h$ to illustrate how the ionized regions follow the distribution of halos, which serve as ionizing sources. At the onset of the reionization process, the Universe was filled mostly with neutral hydrogen atoms, and the corresponding slice plot would appear entirely green. In our case, as we employ the excursion set approach, only ionized and neutral regions are depicted, without capturing fluctuations within the ionized regions. At this stage, reionization is approximately 45\% complete. 
As shown, the 21\,cm signal predominantly arises from regions with fewer low-mass halos, as these areas remain neutral. It is also evident that ionized regions (green) have formed around halos, as they produce sufficient UV photons to generate ionized bubbles. Over time, some of these bubbles merge, forming even larger ionized structures. There are also a few low-mass halos in the neutral regions that have not yet generated enough ionizing photons to form bubbles around them. As redshift decreases, the number density of halos increases, leading to a higher production of ionizing photons. In the final stages of reionization, these ionized bubbles will merge to form a fully ionized Universe.

The cells are projected according to a frequency resolution of \( \delta_\nu = 2.2 \) GHz, meaning this slice accounts for more halos than the 21\,cm signal panel. Additionally, we apply the beam-smearing effect with a full-width half-maximum (FWHM) beam size of \( \theta_{\rm FWHM} = 1 \) arcmin. The brightness of the intensity map is strongly correlated with the astrophysical properties of galaxies, which we model here using the relation between line luminosity and SFR. For a range of models, the intensity of the map can vary by more than a factor of 10 \citep{limpy}. Bright regions in the map trace the overdense regions of the Universe, where high-mass halos have formed. In reality, there is scatter around the mean relationship between line luminosity and SFR, and stochasticity in the star formation process in galaxies can further influence the intensity distribution of the map.

\begin{figure}[h]
\includegraphics[width=0.5\textwidth]{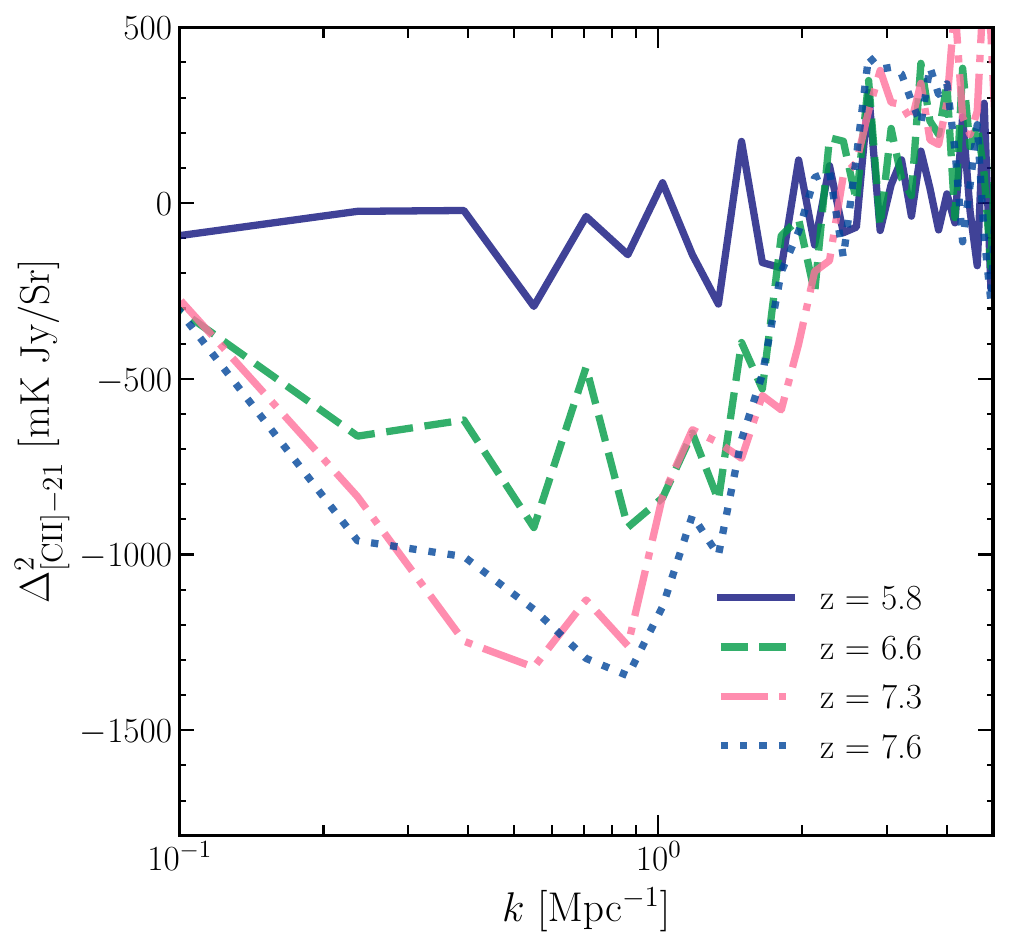} 
\caption{We show the cross-correlated signal between the 21 cm and [CII] line emissions at four different redshifts. The anti-correlation between these two signals reaches its maximum value when the Universe is half ionized and half neutral.}
\label{fig:lim_x_21cm}
\end{figure}

After generating the 21\,cm signal and [CII] intensity grids at the redshifts of our interest, we calculate the three-dimensional power spectrum of these grids. We present our results in terms of the dimensionless power spectrum, $\Delta^2(k) = k^3 P(k)/{2 \pi^2}$, where $P(k)$ is the 3D power spectrum. In Figure \ref{fig:ps_all}, we display the auto power spectra of the 21\,cm and [CII] signals at four redshifts: \( z \sim 7.6 \), 7.3, 6.6, and 5.8. In our case, the amplitude of the 21\,cm signal is highest at \( z \sim 7.6 \), when reionization is approximately 38\% complete, and it gradually decreases as the ionization fraction increases. At \( z \sim 7.6 \), the 21\,cm auto power spectrum is 2.7 times larger than the auto power spectrum at $z \sim 5.8$ at \( k \sim 0.1\,h/\mathrm{Mpc} \) and 7.5 times larger at \( k \sim 1\,h/\mathrm{Mpc} \). Conversely, an opposite trend is observed for the [CII] auto power spectrum at these redshifts. We find that the [CII] signal becomes stronger as the redshift decreases. This is because the number density of halos increases at lower redshifts, and since LIM accounts for the cumulative emission from all sources, the intensity correspondingly increases. For instance, the [CII] power spectrum at \( z\sim 5.8 \) is approximately 9 times stronger at \( k \sim 0.1\,h/\mathrm{Mpc} \) than that at \( z \sim 7.6 \). Moreover, the [CII] power spectrum can vary by more than two orders of magnitude across a range of models \citep{Karoumpis2021}.

The redshift evolution of the cross-correlated signal [CII]-21\,cm is depicted in Figure \ref{fig:lim_x_21cm}. At large scales, these [CII] and 21\,cm signals exhibit anti-correlation, reaching maximum anti-correlation around $z \sim 7.6$, with a signal level similar to $z \sim 7.3$ due to minimal changes in the ionization fraction. As reionization progresses, the degree of anti-correlation diminishes, and the mean cross-correlation approaches zero as reionization nears completion. Both the shape and amplitude of the large-scale anti-correlations are sensitive to $x_e$ and $M_{\rm min}$, where the prominent bump arises from ionized bubbles. Towards the end of reionization, the large-scale bump in the [CII]-21\,cm signal vanishes as ionized bubbles merge, leading to a fully ionized Universe.

\section{Experiments and foreground}\label{sec:exp_forecast}

We delve into the implications of detecting the LIM-21\,cm signal over multiple redshifts by using two experiments. The significance of such a detection would be the direct observation of the EoR through cross-correlation techniques. However, our focus extends beyond mere detection; we aim to determine if this observation can constrain key physical properties of reionization, such as the ionization fraction, the mass of ionizing sources, and other relevant parameters. Furthermore, we explore the possibility of constraining parameters that are intrinsic to the LIM signal itself. By doing so, we can estimate how reionization morphology parameters are influenced by the parameters related to the ISM.

\begin{figure}[h]
\includegraphics[width=0.5\textwidth]{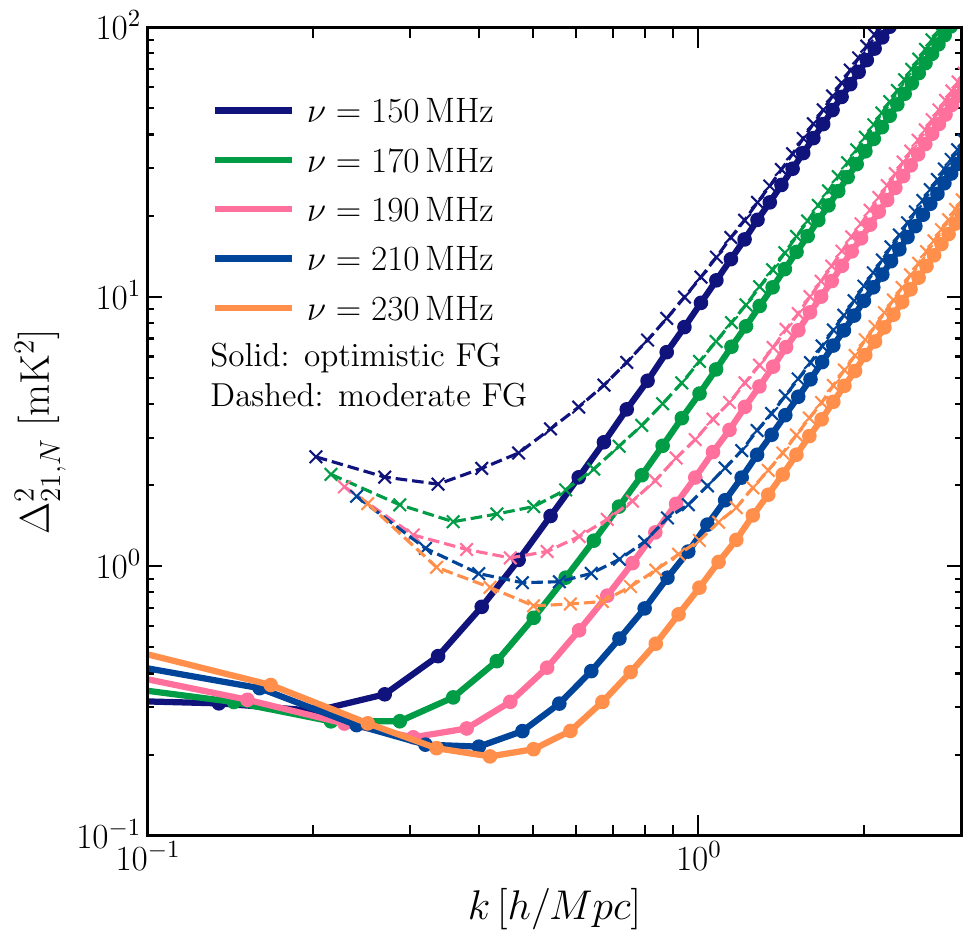} 
\caption{The foreground contamination of the 21\,cm signal coming from the EoR. The solid lines represent the optimistic foreground model, where the dashed lines are the moderate foreground models implemented in 21cmSense package. These models consist of both extragalactic foreground, thermal noise and sample variance. We show the five frequency bands which are necessary for measuring the 21\,cm signal for estimating the LIM-21\,cm cross-correlations.  }
\label{fig:21cmforeground}
\end{figure}

To estimate parameter constraints from the LIM-21 cm cross-correlation signal, we employ Fisher matrix analysis \citep{Dan-Coe-Fisher-Matrix}. The Fisher matrix quantifies the statistical uncertainties in the model parameters and is defined as:

\begin{equation}
F_{ij} = \sum_k \left( \frac{\partial P(k)}{\partial \theta_i} \right)^\dagger \frac{1}{\sigma_k^2} \left( \frac{\partial P(k)}{\partial \theta_j} \right),
\label{eq:Fisher_Matrix}
\end{equation}

where \( \theta_i \) and \( \theta_j \) represent the set of parameters we aim to constrain from the observation of LIM-21\,cm cross-correlation. The term \( \sigma(k) \) denotes the error covariance matrix, which depends on the sensitivity and configuration of a combination of particular 21\,cm and LIM experiments. For the sake of simplicity in this work, we will assume Gaussian errors and a diagonal covariance matrix. However, for highly precise measurements of parameters, one must consider the full covariance matrix, which will be feasible as data quality improves \citep[See][for more details]{fronenberg2024forecasts}.

\begin{figure*}[t]
\centering
\includegraphics[width=1\textwidth]{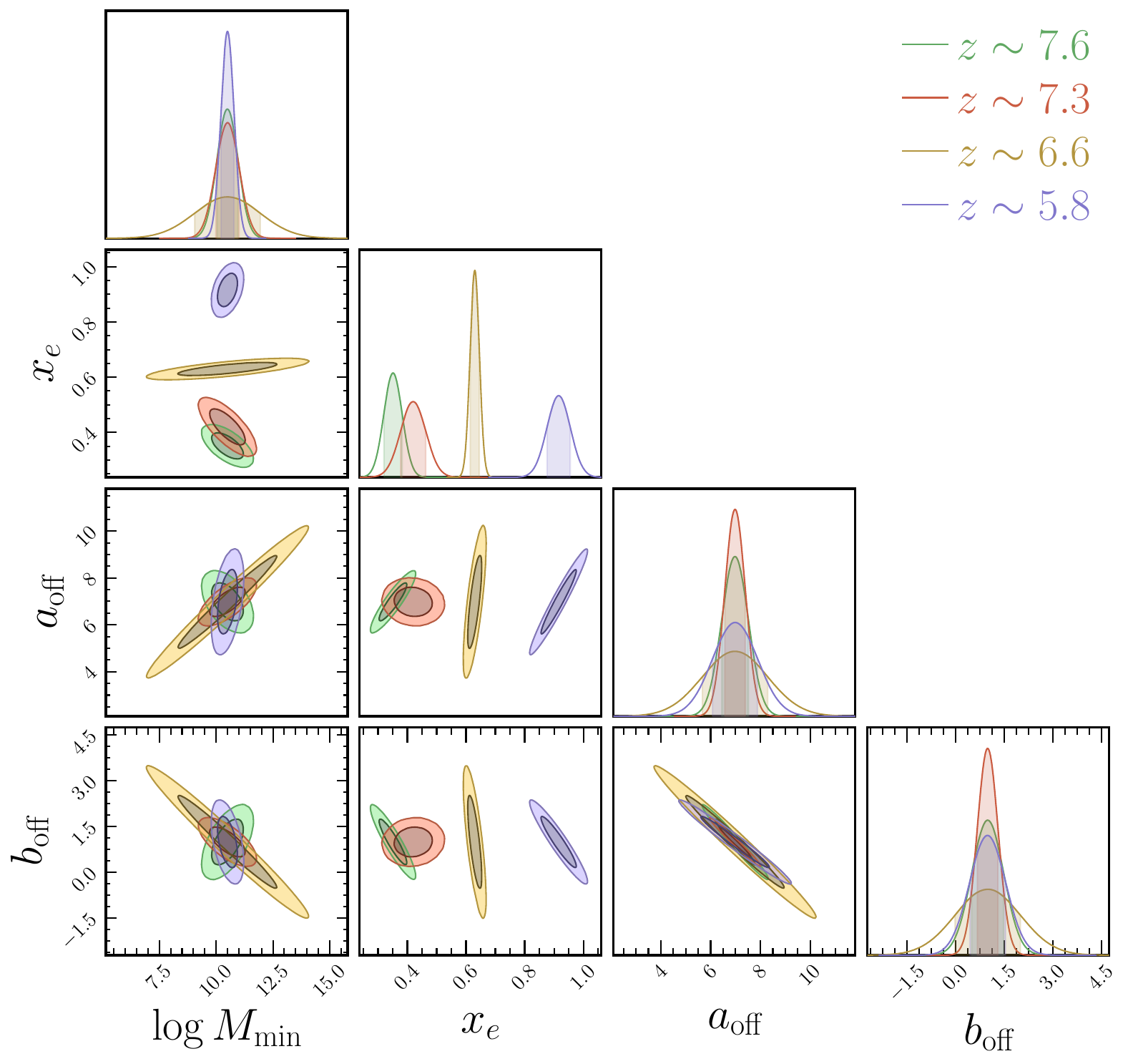} 
\caption{Parameter constraints and their redshift dependencies as inferred from the [CII]-21\,cm cross-correlations. We use SKA-low and FYST-like instruments to forecast the parameters at four redshifts. As the ionization fraction $x_e$ increases at lower redshifts, the contours shift to higher values of $x_e$. This indicates that we can measure the ionization fraction precisely at these redshifts to better understand the morphology of reionization.}
\label{fig:triangle_z}
\end{figure*}

\subsection{Experiments}
In this study, we primarily consider an EoR-Spec-like experiment at FYST for high-redshift [CII] observations \citep{CCAT-prime2021}. The FYST is designed to observe the [CII] line emission, 220 to 410 GHz, which will enable us to map the Universe from z=3.5 to 7.6. For the [C\,\textsc{ii}] intensity-mapping power spectrum, we adopt the
white-noise levels corresponding to the four spectrometer channels \citep{CCAT-prime2021}.
The resulting 3D noise power spectra are
\(P_N^{\rm [CII]} = 2.6\times10^{9}\,{\rm Mpc}^3\,{\rm Jy}^2\,{\rm sr}^{-2}\)
at 220~GHz,
\(4.9\times10^{9}\,{\rm Mpc}^3\,{\rm Jy}^2\,{\rm sr}^{-2}\)
at 280~GHz,
\(3.9\times10^{10}\,{\rm Mpc}^3\,{\rm Jy}^2\,{\rm sr}^{-2}\)
at 350~GHz, and
\(1.2\times10^{11}\,{\rm Mpc}^3\,{\rm Jy}^2\,{\rm sr}^{-2}\)
at 410~GHz. These values set the thermal-noise floor for the
[C\,\textsc{ii}] auto-power spectrum in our forecasts, and we interpolate
the noise power spectrum for intermediate frequency channels that fall
within this range.

For the CO observations, we adopt a strategy similar to that of the CO Mapping Array Project (COMAP)-like experiment \citep{COMAP-Stutzer-2024, COMAP-Breysse-Reionization-2022}. COMAP is designed to map the distribution of CO line emissions, which serve as tracers of molecular gas in galaxies. We extend this strategy by scaling the observations to cover different rotational transitions of CO, specifically CO(1-0) through CO(4-3), across the redshift range of 5 to 8. These transitions provide complementary information about the molecular gas content and the star formation activity in galaxies during this critical epoch.

For our forecasts of second-generation CO line-intensity mapping experiments, we consider a representative setup with a Gaussian beam of full width at half maximum \(\theta_{\rm FWHM} = 4.5'\), a survey area of \(\Omega_{\rm field} = 8~{\rm deg}^2\), and a total on-sky integration time of \(t_{\rm int} = 2000~{\rm hr}\). We assume two observing bands: \(\nu_{\rm obs} = 26\text{--}34~{\rm GHz}\) targeting CO(2–1) and \(\nu_{\rm obs} = 13\text{--}17~{\rm GHz}\) targeting CO(1–0). The instrumental noise power spectrum is computed following a COMAP-like formalism, in which the system temperature \(T_{\rm sys}(\nu_{\rm obs})\) is obtained by interpolating reference values from \citet{Selina2018}, and the 3D noise power for a given line and central observing frequency \(\nu_c\) can be expressed as \citep{Li2016, COMAP-Breysse-Reionization-2022}:
\begin{equation}
    P_N(\nu_c) \;=\;
    \frac{T_{\rm sys}^2(\nu_c)\,\Omega_{\rm field}\,D^2(z)\,c\,(1+z)^2}
         {N_{\rm feeds}\,t_{\rm int}\,H(z)\,\nu_{\rm line}},
\end{equation}
where \(z = \nu_{\rm line}/\nu_c - 1\) is the redshift of the CO transition with rest frequency \(\nu_{\rm line}\), \(D(z)\) is the comoving distance, \(H(z)\) is the Hubble parameter, \(c\) is the speed of light, and \(N_{\rm feeds}=19\) is the number of (single-polarization) feeds. For the fiducial survey configuration described above, we obtain
\begin{equation}
    P_{N,0}^{\rm CO(2\!-\!1)} \;=\; 5.66\times10^{4}\;{\rm Mpc}^3\,\mu{\rm K}^2 .
\end{equation}

\begin{equation}
    P_{N,0}^{\rm CO(1\!-\!0)} \;=\; 2.48\times10^{5}\;{\rm Mpc}^3\,\mu{\rm K}^2 .
\end{equation}
For alternative survey designs, the noise power spectrum can be rescaled as
\begin{equation}
    P_N \;=\; P_{N,0}\;
    \left(\frac{\Omega_{\rm field}}{8~{\rm deg}^2}\right)\;
    \left(\frac{t_{\rm obs}}{2000~{\rm hr}}\right)^{-1},
\end{equation}
where \(P_{N,0}\) denotes the fiducial values quoted above for each CO transition, and \(t_{\rm obs}\) and \(\Omega_{\rm field}\) are the modified total observing time and survey area, respectively.

For the 21\,cm forecasts with SKA1-LOW, we model the instrumental
thermal noise power spectrum as \citep{Parsons2012}
\begin{equation}
    P_N(k,z) \;=\; \frac{T_{\rm sys}^2(\nu_c)}{t_{\rm int}\,\Delta\nu}\;
    \frac{D^2(z)\,\Delta D}{n_b(k_\perp,z)} ,
    \label{eq:PN_SKA}
\end{equation}
where \(T_{\rm sys}\) is the system temperature, \(t_{\rm int}\) is the
total integration time, \(\Delta\nu\) is the frequency bandwidth,
\(D(z)\) is the comoving distance to redshift \(z\), \(\Delta D\) is the
comoving radial depth corresponding to \(\Delta\nu\), and
\(n_b(k_\perp,z)\) is the density of baselines that measure the
transverse Fourier mode \(k_\perp\). We adopt the SKA-Low–like experimental configuration as described in \citet{Dumitru2018} and \citet{Ghara2016}. For SKA1-LOW we adopt \citep{Thompson2017}
\begin{equation}
    T_{\rm sys}(\nu_c) \;=\;
    60~{\rm K}\left(\frac{300~{\rm MHz}}{\nu_c}\right)^{2.25},
    \label{eq:Tsys}
\end{equation}
appropriate for a sky-noise–dominated low-frequency instrument at
mid-latitude, with \(\nu_c\) the central observing frequency. The array
consists of \(N_{\rm ant}=512\) stations, each with effective collecting
area \(A_{\rm eff}=962~{\rm m}^2\), and baselines ranging from
\(b_{\min}=16.8~{\rm m}\) to \(b_{\max}=4.03\times10^{4}~{\rm m}\); these
baselines determine \(n_b(k_\perp,z)\) through the usual mapping between
baseline length \(|\boldsymbol{b}|\) and transverse mode
\(k_\perp = 2\pi|\boldsymbol{b}|/(\lambda D)\), where
\(\lambda = c/\nu_c\). We assume a survey strategy with an observing
time of 6\,hr per day for 120 days, yielding a total integration time
\begin{equation}
    t_{\rm int} = 6~{\rm hr} \times 120 = 720~{\rm hr},
\end{equation}
and a bandwidth of \(\Delta\nu = 6~{\rm MHz}\) around \(\nu_c\). The
corresponding comoving line-of-sight depth is
\begin{equation}
    \Delta D(z) \;=\;
    \left.\frac{{\rm d}D}{{\rm d}\nu}\right|_{\nu_c}\Delta\nu
    \;=\;
    \frac{c\,(1+z)^2}{H(z)\,\nu_{21}}\,\Delta\nu,
\end{equation}
where \(\nu_{21}=1420.4~{\rm MHz}\) is the rest-frame 21\,cm frequency
and \(H(z)\) is the Hubble parameter. Given a cosmology (to compute
\(D(z)\) and \(H(z)\)) and a baseline distribution (to compute
\(n_b(k_\perp,z)\)), Eqs.~(\ref{eq:PN_SKA})–(\ref{eq:Tsys}) specify the
thermal noise power spectrum \(P_N(k,z)\) for SKA1-LOW in units of
\({\rm K}^2({\rm Mpc}/h)^3\).

To maximize the scientific return, we explore different combinations of these observational strategies. We assume that the FYST-like, COMAP-like, SKA-low-like experiments will all survey a common area of 8 square degrees on the sky. This common survey area allows us to combine data from different instruments, enhancing our ability to constrain the physical properties of the early Universe, such as the ionization fraction, the mass of ionizing sources, and the interplay between different components of the interstellar and intergalactic media.

By considering various combinations of these experiments, it will be feasible to develop a comprehensive observational strategy that leverages the strengths of each instrument. This integrated approach will provide a more complete picture of the high-redshift Universe, improving our understanding of galaxy formation, the reionization process, and the evolution of cosmic structures during the EoR.

\begin{figure*}[t]
\centering
\includegraphics[width=1\textwidth]{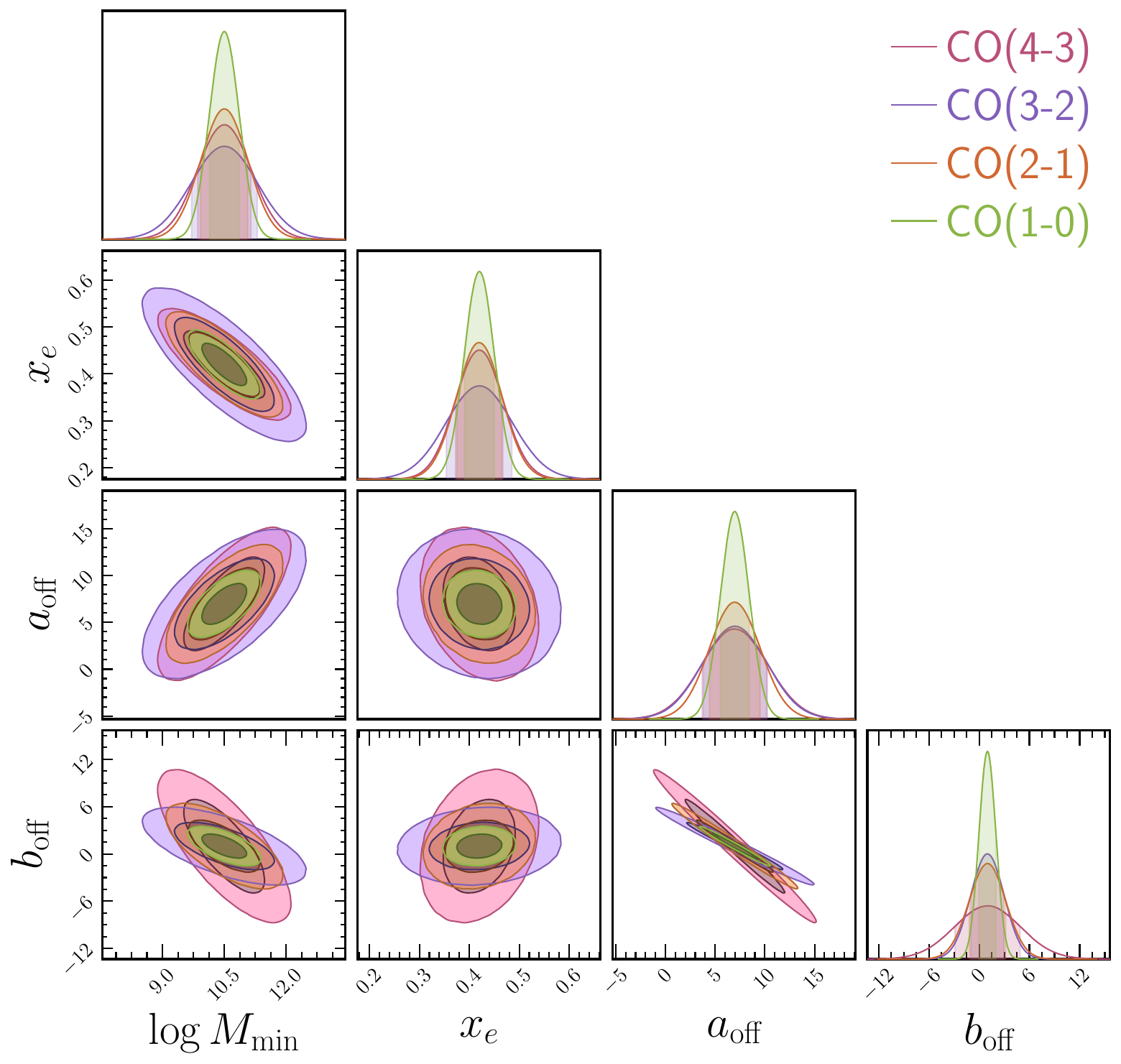} 
\caption{The parameter constraints are determined from LIM–21\,cm cross-correlations, assuming CO(1-0) to CO(4-3) line emissions. We consider the ``optimistic'' ground-based scenario while using SKA-Low as our experiment, and a COMAP-like experiment for measuring different CO transitions. Our forecasts show that CO(1-0) provides the tightest constraints on the parameters; however, a joint analysis would be useful for breaking degeneracies among the parameters.
}
\label{fig:triangle_lines}
\end{figure*}

\subsection{Foregrounds and interlopers}
The removal of foreground contamination from the 21\,cm signal at both the map and power spectrum levels is a challenging task, and a detailed analysis of foregrounds is beyond the scope of this paper. We employed the 21cmSense package to compute the noise power spectrum, including foreground contamination, for various experimental setups. For reference, we primarily use the SKA-low experiment, which is pre-implemented in the package. We set the buffer scale at $k \sim 1$ $h$/Mpc for our analysis.

In Figure~\ref{fig:21cmforeground}, we present the foreground contamination at several frequencies relevant for detecting the 21\,cm signal from the EoR. We show the foreground power spectra for both the ``optimistic'' and ``moderate'' foreground models as implemented in \texttt{21cmSense} \citep{21cmsense1, 21cmsense2}. At $\nu = 150$\,MHz, the moderate foreground power spectrum is $\sim$2.4 times more dominant than the optimistic foreground model. Furthermore, at 230\,MHz, this ratio increases to 3.5. At $k \sim 2$\,h\,Mpc$^{-1}$, the ratio between the foregrounds at 150\,MHz and 230\,MHz is 13.4.

To model contamination from line confusion, we first identify all spectral transitions that overlap with the target emission within a given instrumental frequency channel, defined by the experimental bandwidth $\Delta\nu$. For each channel, we compute the redshift at which every relevant spectral line would be observed at the same frequency as the signal of interest. This procedure determines the full set of foreground and background transitions that act as interlopers. We then construct separate intensity maps for each of these contaminating lines at their corresponding redshifts, ensuring that all emissions sharing the same observed frequency window are consistently included in the analysis. Having identified the relevant interloping lines and their associated redshift ranges, we model their contributions by assigning line luminosities to dark matter halos using empirical relations. For the CO interlopers, we adopt the model of \citet{Kamenetzky2016}, which is the same model used for the target CO signal, but evaluated at the appropriate redshifts corresponding to each interloping transition.

For interloper simulations, we create $100$ slices across a redshift range from $0$ to $20$, with each slice corresponding to an approximate age of $130$ Myr. A single random seed was used to generate snapshots for all $100$ redshifts. This set of snapshots, all evolving from a single Gaussian random initial condition and maintaining the same seed, is designated as ``Sim-set 1''. To capture variations along different lines of sight and intrinsic sample variance, this process was repeated $13$ times with different random seeds. Consequently, we obtained $13$ distinct simulation sets (``Sim-set 1'' to ``Sim-set 13''), each evolving from different Gaussian random fields and providing unique Universe representations \citep[see][for more details]{Roy-Lim-LLX}. These simulations were used to investigate interloper contributions in the detection of [CII]-21\,cm and COs-21\,cm cross-correlation. 

\section{Forecasts}\label{sec:forecast}

In this section, we discuss the impact of parameters related to the reionization process on the LIM$-$21\,cm cross-correlation. We aim to constrain the minimum mass of ionizing sources ($\log M_{\rm min}$), ionization fraction ($x_e$), amplitude ($a_{\rm off}$), and power law index ($b_{\rm off}$) of the line intensity to star formation rate (SFR) relation. Due to the complexity of reionization morphology, we assumed that $\log M_{\rm min}$, $a_{\rm off}$, and $b_{\rm off}$ do not evolve with redshift; only $x_e$ changes with redshift. As the amplitude of the anti-correlation between 21\,cm and LIM changes with redshift, we aimed to investigate if we can reconstruct the complete ionization history of the universe by measuring $x_e$ at different redshift slices during the EoR.

\begin{figure*}[t]
\centering
\includegraphics[width=1\textwidth]{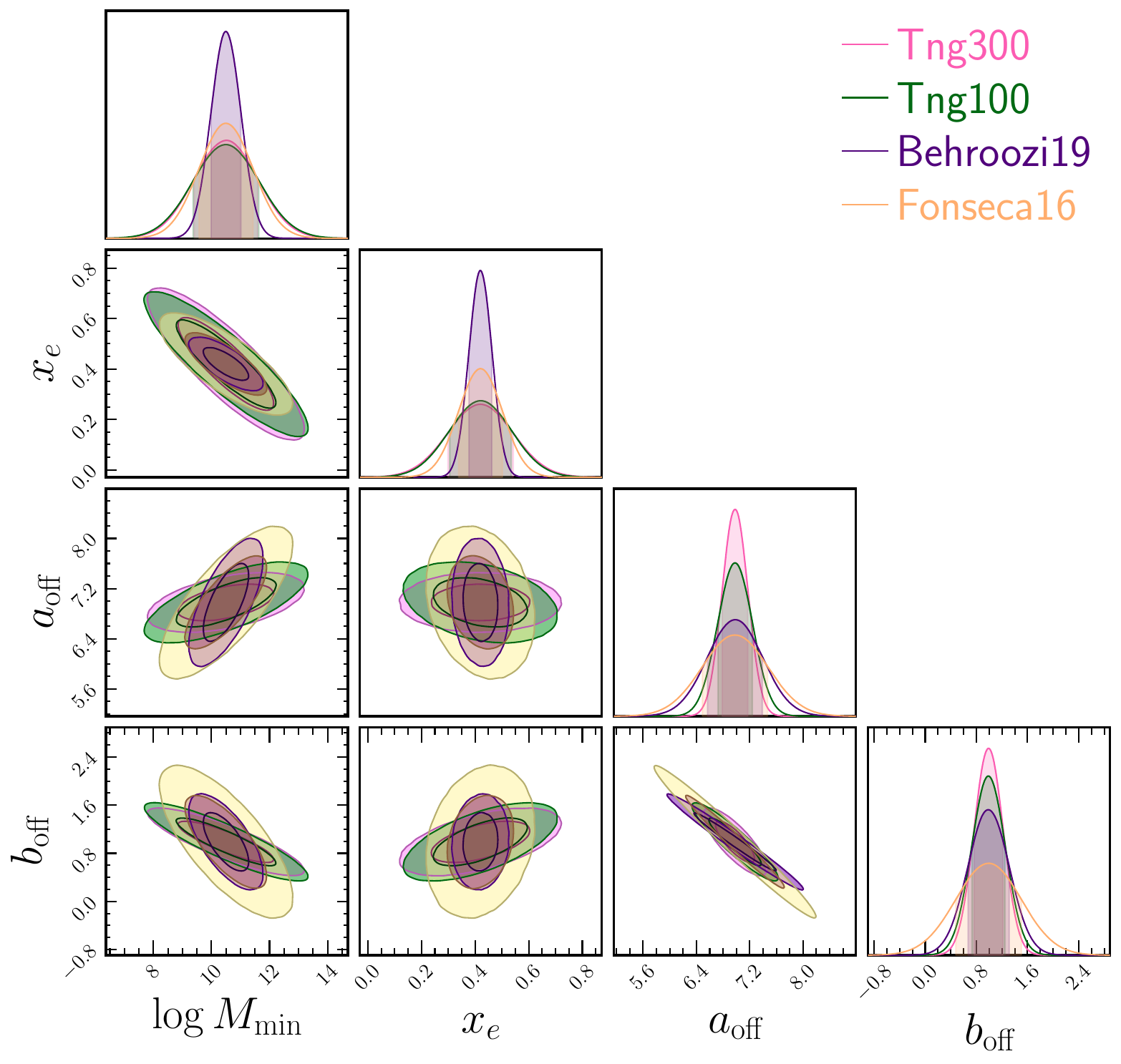} 
\caption{The change in parameter countours if we change the models of star formation history. For our fiducial case, we use ``Behroozi19'' SFR model. This analysis shows that we get the tighter constraints on $M_{\rm min}$, $x_e$, and $b_{\rm off}$ for ``Behroozi19'', while we get tightest constraint on $a_{\rm off}$ for Tng300 SFR model.}
\label{fig:triangle_sfr}
\end{figure*} 

\subsection{Parameter constraints}
In Figure \ref{fig:triangle_z}, we present the parameter constraints inferred from the anti-correlated signal of [CII] and 21\,cm at redshifts 7.6, 7.3, 6.6, and 5.8. The corresponding parameter values are provided in Table \ref{tab:model_params_CII}. For this analysis, we adopt an optimistic foreground scenario for the SKA-low experiment and a FYST-like experiment for measuring the [CII] at overlapping redshifts. This choice of experimental setups ensures that the cross-correlation signal of [CII] and 21\,cm is accurately captured, allowing us to probe the ionization history of the Universe.

Under our chosen parameterization, \( x_e \), is the only parameter that changes across redshift. Meanwhile, we have assumed that the LIM parameters, such as \( a_{\rm off} \) and \( b_{\rm off} \), remain constant across the redshift bins. We can thus use the forecasted constraints on \( x_e \) to predict how well a [CII]-21\,cm measurement can constrain the reionization history. Our results indicate that it is possible to measure the ionization history of the Universe in different redshift bins with statistical significance ranging from 9\( \sigma \) to 40\( \sigma \), enabling us to trace the process of reionization across these epochs. Such precise measurements of the ionization fraction are invaluable in distinguishing between models of early-time and late-time reionization. Furthermore, these measurements can complement observations from other probes, such as the Ly\( \alpha \) and Ly\( \beta \) forest, Quasar near-zone observations, and optical depth constraints from the CMB.

\begin{table}[h]
    \centering 
   \begin{tabular}{ccccc}
        \hline
		Model & $\log M_{\rm min}$ & $x_{e}$ & $a_{\rm off}$ & $b_{\rm off}$ \\ 
		\hline
		$z\sim 7.6$ & $10.5\pm 0.46$ & $0.35\pm 0.031$ & $7.0\pm{0.53}$ & $0.99\pm{0.51}$ \\ 
		$z\sim 7.3$ & $10.5\pm{0.52}$ & $0.42\pm{0.044}$ & $7.0\pm{0.42}$ & $0.99\pm{0.32}$ \\ 
		$z\sim 6.6$ & $10.5\pm 1.41$ & $0.63\pm {0.016}$ & $7.0\pm 1.30$ & $0.99\pm 1.03$ \\ 
		$z\sim 5.8$ & $10.5\pm{0.29}$ & $0.91\pm{0.041}$ & $7.0\pm{0.90}$ & $0.99\pm{0.54}$ \\ 
		\hline
    \end{tabular}
    \caption{Forecasted constraints on the parameters related to reionization and LIM, as inferred from the [CII]-21\,cm cross-correlation, for \(z \sim 5.8\) to \(7.6\).}
\label{tab:model_params_CII}
\end{table}

\begin{table*}
    \centering
    \begin{tabular}{ccccc}
        \hline
		Model & $\log M_{\rm min}$ & $x_{e}$ & $a_{\rm off}$ & $b_{\rm off}$ \\ 
		\hline
		CO(4-3) & $10.5\pm{0.64}$ & $0.42\pm{0.048}$ & $7.0\pm{3.4}$ & $0.99\pm{3.8}$ \\ 
		CO(3-2) & $10.5\pm{0.79}$ & $0.42\pm{0.064}$ & $7.0\pm 3.2$ & $0.99\pm 2.0$ \\ 
		CO(2-1) & $10.5\pm{0.57}$ & $0.42\pm 0.045$ & $7.0\pm{2.6}$ & $0.99\pm{2.1}$ \\ 
		CO(1-0) & $10.5\pm 0.36$ & $0.42\pm{0.030}$ & $7.0\pm{1.4}$ & $0.99\pm 1.0$ \\ 
		\hline
    \end{tabular}
    \caption{Parameter constraints at $z\sim 7.3$ extracted from the cross-correlation between different CO line emissions and the 21\,cm signal, as described in Figure \ref{fig:triangle_lines}.}
    \label{tab:model_params_co_lines}
\end{table*}

In addition to ionization history, the measurement of parameters associated with the [CII] emission can provide crucial insights into the ionizing sources responsible for reionization. In our analysis, we assume that galaxies are the primary contributors to reionization. We find that the parameter \( a_{\rm off} \) could be measured with a precision of 17\( \sigma \) at \( z \sim 7.3 \) and 5\( \sigma \) at \( z \sim 6.6 \). We also observe strong degeneracies between \( a_{\rm off} \) and \( b_{\rm off} \), which affect the measurement of \( b_{\rm off} \). These degeneracies result in a lower measurement precision for \( b_{\rm off} \), which could be constrained to between 1\( \sigma \) and 3\( \sigma \) at the aforementioned redshifts. 

\begin{table*}
    \centering
  \begin{tabular}{ccccc}
        \hline
		Model & $\log M_{\rm min}$ & $x_{e}$ & $a_{\rm off}$ & $b_{\rm off}$ \\ 
		\hline
		Tng300 & $10.5\pm 1.1$ & $0.42\pm 0.12$ & $7.0\pm{0.20}$ & $0.99\pm{0.23}$ \\ 
		Tng100 & $10.5\pm 1.1$ & $0.42\pm{0.12}$ & $7.0\pm 0.26$ & $0.99\pm 0.26$ \\ 
		Behroozi19 & $10.5\pm 0.51$ & $0.42\pm{0.042}$ & $7.0\pm{0.42}$ & $0.99\pm 0.32$ \\ 
		Fonseca16 & $10.5\pm{0.93}$ & $0.42\pm 0.082$ & $7.0\pm{0.48}$ & $0.99\pm{0.52}$ \\ 
		\hline
    \end{tabular}
        \caption{Summary of the best-fit parameters estimated from the [CII]–21\,cm signal for different SFR models at $z\sim 7.3$, as described in Figure\,\ref{fig:triangle_sfr}.}
    \label{tab:model_params_sfr_model}
\end{table*}

There is another possibility to cross-correlate different CO J-level transitions with the 21\,cm signal. It is expected that these LIM signals would also be anti-correlated at large scales, similar to the [CII] emissions. In principle, low-frequency LIM experiments would be able to cross-correlate CO(1-0) to CO(4-3) line emissions with the 21\,cm signal. However, as the experimental designs for these cross-correlations are not currently available, we assume a modified COMAP-like architecture and scale the noise level for other frequencies that are relevant for higher CO J-level transitions.

In Figure \ref{fig:triangle_lines}, we show the parameter constraints inferred from the cross-correlations between the 21\,cm signal and CO(1-0) to CO(4-3) line emissions. These constraints are presented for a redshift of $z \sim 7.3$, when reionization is approximately 42\% complete. The parameter values are reported in Table \ref{tab:model_params_co_lines}. We find that the cross-correlation with CO(1-0) and 21\,cm yields tighter constraints than those obtained from the other CO transitions, placing a 46\% tighter constraint on the ionization fraction \(x_e\) at this redshift compared to the [CII]-21\,cm cross-correlation discussed earlier. However, the best constraints on the parameters \(a_{\rm off}\) and \(b_{\rm off}\) are obtained from the [CII]-21\,cm cross-correlation, rather than the cross-correlation with CO lines. This is primarily because the [CII] lines are brighter than the CO lines at those redshifts of our interest. 

Among the CO lines, the 21\,cm cross-correlation with CO(1-0) provides the strongest constraint on \(a_{\rm off}\), with a \(5\sigma\) measurement. This constraint decreases to \(2\sigma\) when cross-correlating with CO(4-3) lines. Additionally, the \(b_{\rm off}\) parameter can be constrained at \(1\sigma\) using CO(1-0) and at \(0.3\sigma\) using CO(4-3). Furthermore, all of the CO lines are able to constrain \(x_e\) with higher precision, yielding constraints between \(3.5\sigma\) and \(14\sigma\). This is due to the fact that the cross-correlated signal is primarily sensitive to the ionization fraction \(x_e\) rather than to the LIM parameters \(a_{\rm off}\) and \(b_{\rm off}\). For a more robust constraint, it may be beneficial to combine all of these probes, as this would reduce the uncertainties on the parameters due to the stochasticity in the signal. Using a combination of the [CII]-21\,cm and CO(1-0) to CO(4-3)-21\,cm cross-correlations would allow us to leverage the strengths of each line transition and improve our understanding of the reionization process.

As mentioned earlier, although we kept the line luminosity to SFR relation consistent across the EoR, the overall luminosity of the lines changes due to the variation in the SFR model. Different SFR models map the SFR to the \( M_{\rm halo} \) relation at these redshifts, and thus the choice of SFR model can influence the parameter constraints. To understand the impact of the SFR on the parameter constraints, we computed the LIM-21\,cm cross-correlations using different SFR models and investigated the resulting parameter constraints for each.

In Figure \ref{fig:triangle_sfr}, we illustrate how the parameter constraints change for different SFR models when we use the [CII] lines as the LIM tracer at \( z \sim 7.3 \). The summarized results are reported in Table \ref{tab:model_params_sfr_model}. To generate the LIM maps, we employed \texttt{LIMpy}, varying several SFR models, such as Fonseca16 \citep{{Fonseca:2016}}, TNG100, TNG300 \citep{TNG-gen}, and Behroozi19 \citep{Behroozi2019}. Our analysis reveals that the tightest constraints on \( \log M_{\rm min} \) and \( x_e \) come from the Behroozi19 SFR model. Meanwhile, the TNG300 SFR model provides the tightest constraints on the parameters \( a_{\rm off} \) and \( b_{\rm off} \). This highlights the influence of different SFR models on the derived parameter constraints, with some models favoring tighter constraints on specific parameters, potentially due to variations in the SFR-to-halo mass relationship at the relevant redshifts.

\section{Conclusion}\label{sec:conclusion}
Current and future experiments hold the potential to significantly improve our understanding of the reionization history and the properties of the ionizing sources. By leveraging both 21\,cm and LIM observations, we can gain direct and complementary insights into the evolution of the Universe across different redshifts. However, despite significant theoretical progress, \textit{direct detection} of the 21\,cm signal from the EoR remains elusive, while LIM surveys face challenges in disentangling faint line emission from foregrounds and instrumental systematics.
One exciting avenue for future research is the exploration of cross-correlations between these signals. Such cross-correlation studies will provide a more comprehensive view of the reionization process, as they combine the strengths of both techniques. While several attempts have been made to forecast the detectability of the [CII]-21\,cm signal, a more thorough treatment is necessary to fully account for various factors that could affect the measurements. These include foreground contamination on the 21\,cm signal, interlopers in LIM observations, instrumental noise, and the beam effects from the experimental setup. All of these factors need to be carefully modeled and mitigated in future studies to ensure the reliability and accuracy of the results.

In this work, we apply physically motivated models utilizing the excursion set approach for ionized/neutral regions and the \texttt{LIMpy} package for line intensity mapping based on empirical $\log L_{\rm line}$--$\log(\mathrm{SFR})$ scaling relations to generate [CII]-21\,cm and CO-21\,cm cross-correlation forecasts across four redshifts during the EoR. Our analysis demonstrates that these cross-correlations can constrain the ionization fraction $x_e$ with 9--40$\sigma$ significance across $z = 5.8$--7.6, enabling tomographic reconstruction of reionization history, with CO(1-0)--21\,cm yielding 46\% tighter constraints on $x_e$ compared to [CII]-21\,cm at $z \sim 7.3$. We simultaneously constrain the minimum halo mass of ionizing sources ($\log M_{\rm min}$) to within $\pm 0.3$--1.4\,dex and the line luminosity--SFR scaling parameters $a_{\rm off}$ and $b_{\rm off}$ with 1--17$\sigma$ precision, achieving 17$\sigma$ accuracy for $a_{\rm off}$ at $z \sim 7.3$, with constraint precision strongly dependent on the adopted SFR model---Behroozi19 provides the best recovery of $\log M_{\rm min}$ and $x_e$, while TNG300 yields tightest constraints on luminosity scaling parameters. These results establish that next-generation SKA-low and FYST-like experiments will provide the first direct observational framework to distinguish between early versus late reionization scenarios while simultaneously characterizing the properties of ionizing galaxy populations through multi-tracer intensity mapping cross-correlations.

Besides the great potential of the LIM-21\,cm signal, there are several caveats in our analysis. The excursion set approach to model the 21\,cm field is faster than hydrodynamical simulations; however, this model has several shortcomings, such as photon non-conservation and small-scale artifacts \citep{Choudhury-photon-noncon}. Additionally, the parameters $a_{\rm off}$, $b_{\rm off}$, and $M_{\rm min}$ should depend on redshift, which we have ignored in our forecast. Furthermore, our forecast is based on a simple Fisher analysis, but assuming a non-Gaussian likelihood, as shown in \citet{fronenberg2024forecasts}, could alter the results. Moreover, we account for simple foreground scenarios that are ``optimistic" and ``moderate" in nature, but there could be significant systematics that would increase the errors on parameter constraints. In conclusion, the LIM-21\,cm signal is promising and holds great potential for uncovering details of the reionization epoch; however, its fidelity in determining parameters depends on the foreground contamination of the 21\,cm signal and the presence of interlopers.

\acknowledgments
We thank Nick Battaglia, Hannah Fronenberg, and David Spergel for their helpful discussions. We especially thank Girish Kulkarni for providing the excursion set code used to generate the ionization field. AR acknowledges support from NASA under award number 80NSSC18K1014939. ARP was supported by NASA under award numbers 80NSSC18K1014, NNH17ZDA001N, and
80NSSC22K0666, and by the NSF under award number
2108411. The Flatiron Institute is supported by the Simons Foundation.

\bibliographystyle{aasjournal}
\bibliography{citation}

\end{document}